\documentclass{kapproc} % Computer Modern font calls

\usepackage{procps} 
\input{psfig}

\usepackage[dvips]{graphicx}

\upperandlowercase

\kluwerbib % will produce this kind of bibliography entry:

\begin{document}

\articletitle[Void Hierarchy]
{Void Hierarchy and Cosmic Structure\\}

\author{Rien van de Weygaert$^1$ \& Ravi Sheth$^{2,3}$ }

%% affil, email, and abstract are optional
\affil{$^1$Kapteyn Institute, Univ. Groningen, P.O. Box 800, 9700 AV Groningen, 
The Netherlands \\
$^2$Dept. of Physics and Astronomy, University of Pittsburgh, 
3941 O'Hara St., PA 15260, U.S.A.\\
$^3$ NASA/Fermilab Astrophysics Group, Batavia, IL 60510-0500, U.S.A.\\}
\email{rks12@pitt.edu, weygaert@astro.rug.nl}

%% optional, to supply a shorter version of the title for the running head:
%%\chaptitlerunninghead{}

\anxx{Van de Weygaert \& Sheth}

\begin{abstract}
Within the context of hierarchical scenarios of gravitational structure 
formation we describe how an evolving hierarchy of voids evolves 
on the basis of {\em two} processes, the {\em void-in-void} process and 
the {\it void-in-cloud} process. The related analytical formulation 
in terms of a {\it two-barrier} excursion problem leads to a 
self-similarly evolving peaked void size distribution.
\end{abstract}

%\begin{keywords}
%Large-scale structure of the Universe
%\end{keywords}

\section{Introduction: Excursions}
\noindent Hierarchical scenarios of structure formation have been very succesfull 
in understanding the formation histories of gravitationally bound 
virialized haloes. Particularly compelling has been the formulation of a formalism 
in which the collapse and virialization of overdense dark matter halos within the 
context of hierarchical clustering can be treated on a fully analytical basis. This 
approach was originally proposed by Press \& Schechter (1974), which 
found a particularly useful and versatile formulation and modification in the 
the {\it excursion set formalism} (Bond et al. 1991). 

It is based on the assumption that for a structure to reach a particular nonlinear 
evolutionary stage, such as complete gravitational collapse, the sole condition 
is that its {\em linearly extrapolated primordial density} should attain a certain 
value. The canonic example is that of a spherical tophat overdensity collapsing once 
it reaches the collapse barrier $\delta_c\approx 1.69$. The successive contributions 
to the local density by perturbations on a (mass) resolution scale $S_m$ may 
be represented in terms of a density perturbation random walk, the 
cumulative of all density fluctuations at a resolution scale smaller 
than $S_m$. By identifying the largest scale at which the density passes 
through the barrier $\delta_c$ it is possible (1) to infer at any cosmic epoch the 
mass spectrum of collapsed halos and (2) to reconstruct the merging history of 
each halo (see Fig.~\ref{excur}, top right). 

In this study we demonstrate that also the formation and evolution of foamlike patterns 
as a result of the gravitational growth of primordial density perturbations is liable 
to a succesfull description by the excursion set analysis. This is accomplished by 
resorting to a complementary view of clustering evolution in which we 
focus on the evolution of the {\em voids} in the Megaparsec galaxy and matter 
distribution, spatially {\it th\'e} dominant component. An extensive description 
of this work can be found in Sheth \& Van de Weygaert (2003, MNRAS, submitted).

\begin{figure}[t]
\vskip -3.0cm
\centering\mbox{\hskip -0.5truecm\psfig{figure=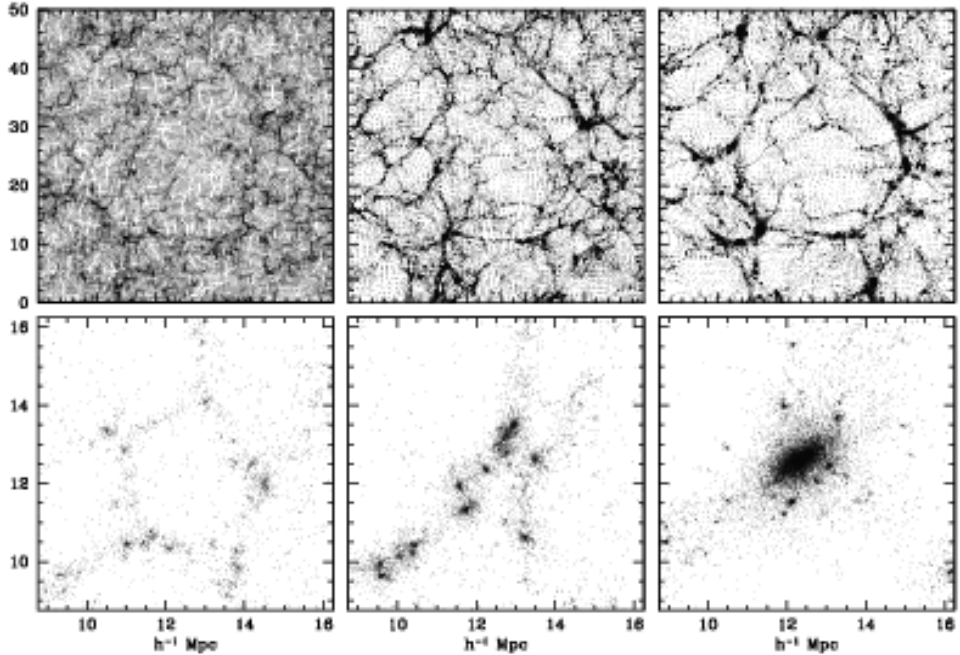,width=12.9cm}}
\vskip -1.75cm
\caption{Illustration of the two essential ``void hierarchy modes'': (top) the 
{\em void-in-void} process (top), with a void growing through the 
merging of two or more subvoids; (bottom) the {\em void-in-cloud} process:  
a void demolished through the gravitational collapse of embedding region.}
\vskip -0.7truecm
\label{voidproc}
\end{figure}
\section{Void Evolution: the two processes}
\noindent Primordial underdensities are the progenitors of voids. Because underdensities 
are regions of suppressed gravitational attraction, representing an effective repulsive 
gravity, matter flows out of their interior and moves outward to the edges of these 
expanding {\em voids}. Voids {\em expand}, become increasingly {\em empty} and 
develop an increasingly {\em spherical} shape (Icke 1984). Matter 
from the void's interior piles up near the edge: usually a ridge forms 
around the void's rim and at a characteristic moment the void's interior 
shells take over the outer ones. At this {\em shell-crossing} epoch the void 
reaches maturity and becomes a nonlinear object expanding self-similarly, the 
implication being that the majority of observed voids is at or near this stage  
(Blumenthal et al. 1992). As voids develop from underdensities 
in the primordial cosmic density field, the interaction with internal substructure and 
external surrounding structures translates into a continuing process of 
hierarchical void evolution (Dubinski et al. 1993, Van de Weygaert \& 
Van Kampen 1993). 

\begin{figure}[h]
\vskip -0.8truecm
\centering\mbox{\hskip -7.0truecm\psfig{figure=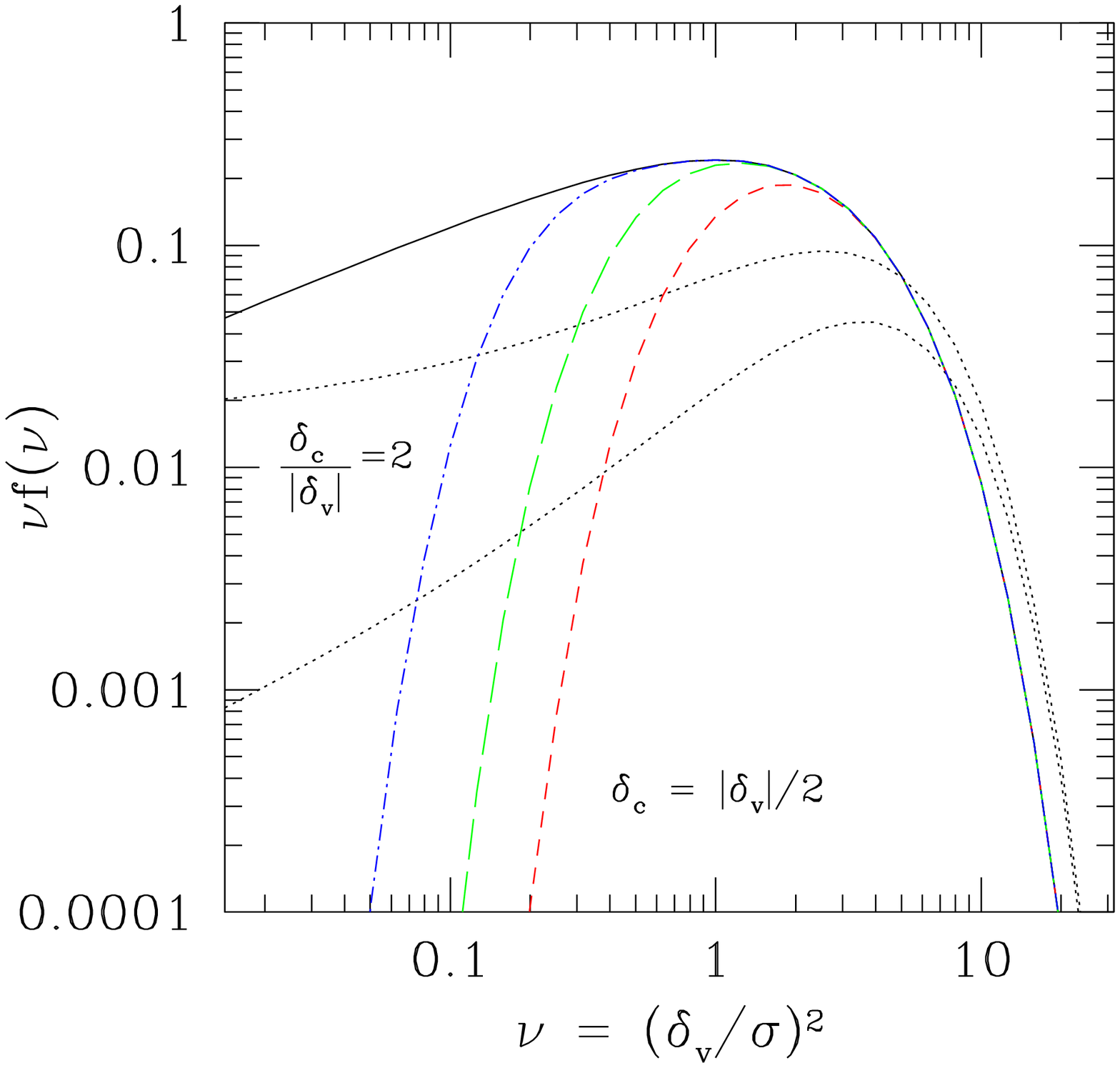,height=8.0cm}}
\vskip -9.5cm
\centering\mbox{\hskip 6.3truecm\psfig{figure=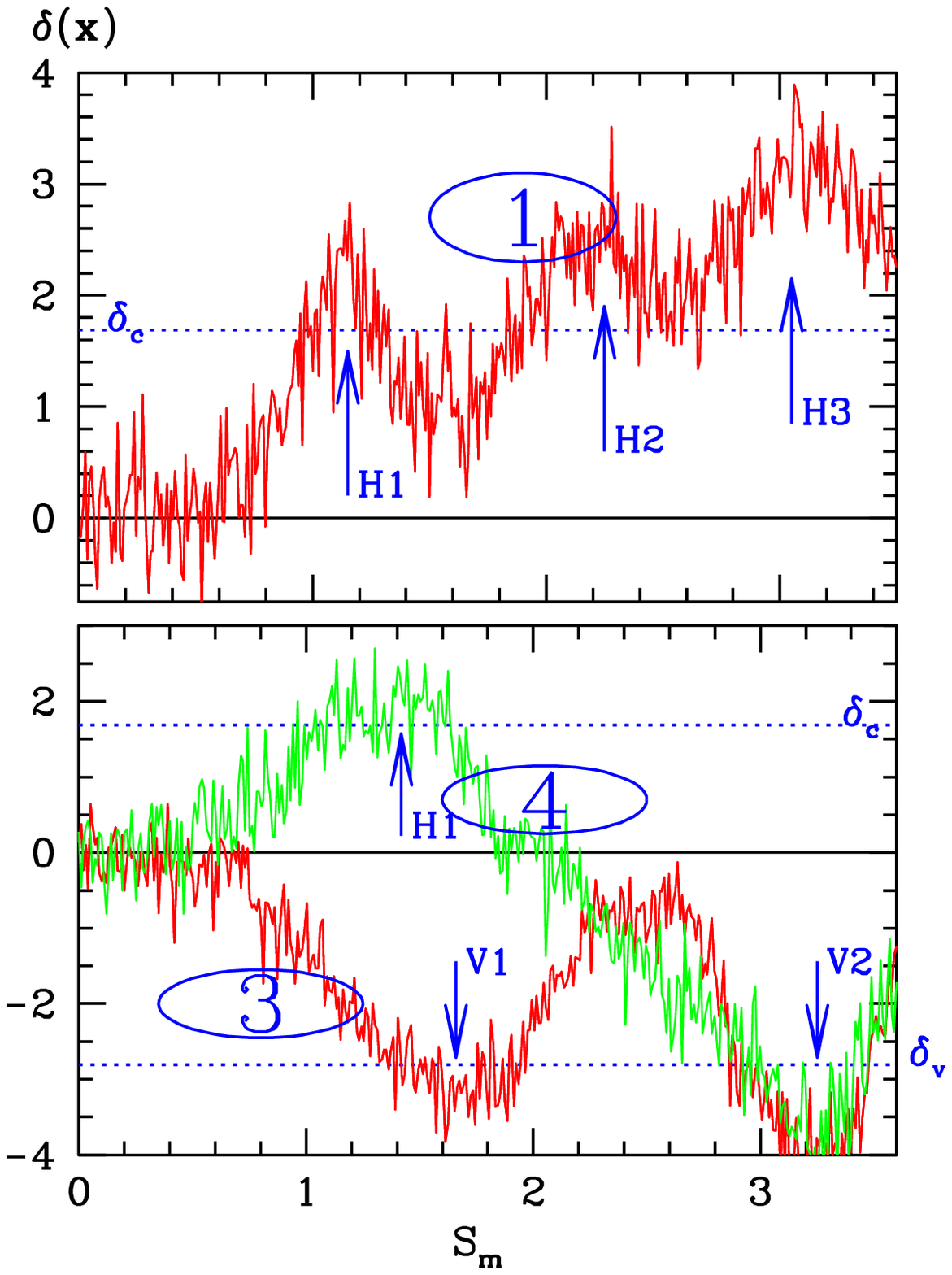,height=10.0cm}}
\vskip -0.75cm
\caption{{\bf Left}: Void Size Probability Function for various values of the 
{\em void-and-cloud} parameter $\delta_c/\delta_v$: 1/2 (short dashed), 1 (dashed), 
2 (dot-dashed), and $\infty$ (solid). For reference, the dotted curves 
are predictions by the ``adaptive peaks model'' of Appel \& Jones (1990). 
{\bf Right}: Excursion Set Random Walks, local density $\delta({\bf x})$ as function 
of mass resolution $S_m$. Clearly, the fluctuations are larger as resolution $S_m$ 
increases. (Top) Halo formation. Horizontal dashed bar is the collapse barrier $\delta_c$. (Bottom) Fate of void V2, illustrated by 2 possible 
random walks: curve ``4'' is a {\em void-in-cloud} process, curve ``3'' a 
{\em void-in-void} process.}
\vskip -1.1truecm
\label{excur}
\end{figure}

The evolution of voids resembles that of dark halos in that large voids form from 
mergers of smaller voids that have matured at an earlier cosmic time 
(Fig.~\ref{voidproc}, top row). However, in contrast to dark halos, the fate of voids 
is ruled by two processes. Crucial is the realization that the evolving void hierarchy 
does not only involve the {\em void-in-void} process but also an additional aspect, the 
{\em void-in-cloud} process. Small voids may not only merge into larger encompassing 
underdensities, they may also disappear through collapse when embedded within a larger 
scale overdensity (Fig.~\ref{voidproc}, bottom row). In terms of the 
{\em excursion set approach}, 
it means that the {\em one-barrier} problem for the halo population has to be extended 
to a more complex {\em two-barrier problem}. Voids not only should ascertain 
themselves of having decreased their density below the {\em void barrier} $\delta_v$ of 
the {\em shell-crossing} transition. For their survival and sheer existence it is 
crucial that they take into account whether they are not situated within a collapsing 
overdensity on a larger scale which crossed the collapse barrier $\delta_c$. They 
should follow a random walk path like type ``3'' in Fig.~\ref{excur} (bottom right), rather than 
the {\em void-in-cloud} path ``4''. The repercussions of this are far-reaching and 
it leads to a major modification of the void properties and distribution.

\section{Void Distribution: Universal, Peaked and Self-Similar}
\noindent Analytically, the resulting expression follows by evaluating the 
fraction of walks which first cross $\delta_{\rm v}$ at $S$, and which do not 
cross $\delta_{\rm c}$ until after they have crossed $\delta_{\rm v}$ 
(Sheth \& Van de Weygaert 2003). An insightful approximation of 
this distribution, in terms of the ``self-similar'' void density 
$\nu\,\equiv\,\delta_{\rm v}^2/S\,$ is 
\begin{equation}
 \nu f(\nu) \approx \sqrt{\nu\over 2\pi}\,
 \exp\left(-{\nu\over 2}\right)\,
 \exp\left(-{|\delta_{\rm v}|\over\delta_{\rm c}}\,
 {D^2\over 4\nu}-2{D^4\over\nu^2} \right)\,,
 \label{vfvapprox}
\end{equation}
\noindent in which the $D\equiv (|\delta_{\rm v}|/(\delta_{\rm c}-\delta_{\rm v})$ 
parameterizes the the relative impact of void and halo evolution on the hierarchically 
evolving population of voids. 

The resulting distributions, for various values of $D$, are shown in 
Figure~\ref{excur} (left). The void size distribution is clearly {\em peaked about a 
characteristic size}: the {\em void-in-cloud} mechanism is responsible for the demise of 
a sizeable population of small voids. The halo mass distribution diverges towards small 
scale masses, so that in terms of numbers the halo population is dominated by 
small mass objects. The void population, on the other hand, is ``void'' of small voids and 
has a sharp small-scale cut-off. 

Four additional major observations readily follow from this analysis: 
($\bullet$) At any one cosmic epoch we may identify a 
{\em characteristic void size} which also explains why in the present-day foamlike 
spatial galaxy distribution voids of $\sim 20-30h^{-1}\hbox{Mpc}$ are the predominant 
feature; ($\bullet$) The {\em void distribution evolves self-similarly} and the 
{\em characteristic void size increases with time}: the larger voids 
present at late times formed from mergers of smaller voids which constituted the 
dominating features at earlier epochs (Fig.~\ref{voidproc}, top panels); 
($\bullet$) Volume integration shows that at any given time {\em the population of voids 
approximately fill space}, apparently squeezing the migrating high-density matter in 
between them; ($\bullet$) As the size of the major share of voids will be in the order 
of that of the characteristic void size this observation implies that the {\it cosmic matter 
distribution} resembles a {\it foamlike packing of spherical voids of approximately similar 
size and excess expansion rate}. 

In all, a slight extension and elaboration on the original extension formulism enables 
the framing of an analytical theory explaining how the characteristic observed weblike 
Megaparsec scale galaxy distribution, and the equivalent frothy spatial matter 
distribution seen to form in computer simulations of cosmic structure formation, are 
natural products of a hierarchical process of gravitational clustering.
A continuously evolving hierarchy of voids produces a dynamical foamlike pattern whose 
characteristic dimension grows continuously along with the evolution of cosmic 
structure, a Universe which at any one cosmic epoch is filled with bubbles whose 
size corresponds to the scale just reaching {\em maturity}. 
%\begin{chapthebibliography}{1}
%\bibitem[1990]{appjon90} Appel L., Jones B. J. T., 1990, MNRAS, 245, 522
%\bibitem[1992]{bdglp92} Blumenthal G.R., Da Costa L., Goldwirth D.S., Lecar M., 
%Piran T., 1992, ApJ, 388, 234
%\bibitem[1991]{bcek91} Bond J. R., Cole S., Efstathiou G., Kaiser N., 1991, 
%ApJ, 379, 440
%\bibitem[1993]{ddglp93} Dubinski J., Da Costa L.N., Goldwirth D.S., Lecar M., Piran T., 19%93, ApJ, 410, 458
%\bibitem[1984]{vi84} Icke V., 1984, MNRAS, 206, 1
%\bibitem[1974]{ps74} Press W., Schechter P., 1974, ApJ, 187, 425
%\bibitem[2003]{shvdw2003} Sheth R., van de Weygaert R., 2003, MNRAS, submitted
%\bibitem[1993]{vdwvk93} Van de Weygaert R., Van Kampen E., 1993, MNRAS, 263, 481
%\end{chapthebibliography}

\end{document}